\documentclass[conference]{IEEEtran}
\IEEEoverridecommandlockouts
\usepackage{cite}
\usepackage{amsmath,amssymb,amsfonts}
\usepackage{algorithmic}
\usepackage{textcomp}
\usepackage{xcolor}
\usepackage{float}
\usepackage{graphicx, caption, subcaption}
\def\BibTeX{{\rm B\kern-.05em{\sc i\kern-.025em b}\kern-.08em
    T\kern-.1667em\lower.7ex\hbox{E}\kern-.125emX}}
\begin{document}

\title{CSI-Based Localization with CNNs Exploiting Phase Information\\

}

\author{\IEEEauthorblockN{Anastasios Foliadis\IEEEauthorrefmark{1}\IEEEauthorrefmark{2}, Mario H. Casta\~{n}eda Garcia\IEEEauthorrefmark{1}, Richard A. Stirling-Gallacher\IEEEauthorrefmark{1},  Reiner S. Thom\"a\IEEEauthorrefmark{2}}
\IEEEauthorblockA{\IEEEauthorrefmark{1}\textit{Munich Research Center}, \textit{Huawei Technologies Duesseldorf GmbH}, 
Munich, Germany \\
\textit{\IEEEauthorrefmark{2}Electronic Measurements and Signal Processing}, \textit{Technische Universit\"at Ilmenau}, Ilmenau, Germany\\
\{anastasios.foliadis, mario.castaneda, richard.sg\}@huawei.com, reiner.thomae@tu-ilmenau.de}}

\maketitle

\begin{abstract}
In this paper we study the use of the Channel State Information (CSI) as fingerprint inputs of a Convolutional Neural Network (CNN) for localization. We examine whether the CSI can be used as a distinct fingerprint corresponding to a single position by considering the inconsistencies with its raw phase that cause the CSI to be unreliable. We propose two methods to produce reliable fingerprints including the phase information. Furthermore, we examine the structure of the CNN and more specifically the impact of pooling on the positioning performance, and show that pooling over the subcarriers can be more beneficial than over the antennas.
\end{abstract}


\section{Introduction}
Advances in mobile communications and the development of Internet of Things (IoT) has introduced a large variety of new applications in a number of different areas of modern life. One important requirement in several of these applications is the estimation of the user’s position. Although the ubiquity of Global Positioning System (GPS) provides a great solution for outdoor localization, other alternatives are needed indoors. 

Many different solutions have been proposed in the literature for indoor positioning, ranging from classical approaches, like angle of arrival (AoA) and time of arrival (ToA) based, to pattern recognition approaches. More specifically, the ability to store and transmit large amounts of data has directed the focus on using deep learning. Additionally, with the 5th Generation (5G) network being deployed, providing high data rates and bandwidth, the number of antennas on devices is increasing, enabled by the mm-Wave operation frequency. 

For coherent communication, the multi-antenna channel between a user and the base station (BS) is estimated using pilot symbols. The estimated channel referred as channel state information (CSI) can serve as a fingerprint for localization. 

For localization based on fingerprint inputs, a database for a given environment is created offline and during the online phase the UE’s position is estimated, by matching its signal to the fingerprint map. There exist a number of different approaches to implement the mapping. These range from conventional, like maximum likelihood and least squares, to machine learning, like k-nearest neighbors and neural networks.  What is considered a fingerprint also differs depending on the application.
% Received Signal Strength (RSS) has been considered but its performance suffers in complex environments so the focus has recently been on CSI as fingerprint \cite{rssitoCsi}. 

In \cite{widmaier2019practical} the mapping is done using convolutional neural networks (CNNs), achieving a sub-meter accuracy with simulated and real measurements by utilizing the real and imaginary parts of the CSI. In \cite{bast2019csibased}, again a CNN was used with real measurements, but with its inputs consisting of a combination of raw features (real and imaginary), polar features and time-domain features. More complex neural network configurations were used in \cite{Niitsoo2019ADL} using as input the time-domain channel impulse response. The authors of \cite{SudparisEnsemble} were able to achieve sub-centimeter accuracy by employing a denoising technique and an ensemble of neural networks. They considered only the magnitude of the channel, since they identified that phase measurements at the same position can change over time.

Phase spatial inconsistency arising because of implementation aspects is a common issue affecting the CSI. For this reason several prior works either ignore or completely reject the phase and focus primarily on the magnitude. However, the phase could embed important information of the underlying channel for localization purposes. In \cite{wangPhaseCalibration} and \cite{monalisa}, a transformation per antenna is proposed to calibrate the phase of multi-antenna measurements used as inputs to a CNN.

%In this paper, we focus on the proper utilization of the phase. For this purpose, we propose two techniques to make the phase measurements consistent and reliable for localization. We employ CNNs using the both the magnitude and the processed phase of the CSI and show that these techniques improve the localization accuracy. Additionally we try to optimize the structure of the CNN for this particular problem by examining the impact its pooling layer has on the localization performance. 
 
In this paper, we propose two techniques to obtain a processed phase which is consistent and reliable for localization. In contrast to \cite{wangPhaseCalibration} and \cite{monalisa}, we propose the same transformation across all antennas to preserve valuable AoA information. By employing CNNs, we show that the use of the processed phase improves the localization accuracy compared to when using the raw phase. To optimize the CNN based on the structure of the CSI, we also examine the impact of the pooling layer on the localization performance.

In the remainder of this paper we describe the system model in Section II. In Section III, we present the proposed methods to address the phase issues. In Section IV we introduce the machine leaning approach that we utilize and in Section V we present the results of our simulations. We present our conclusions in Section VI.

\section{System Model and Database Description}

\subsection{System Model}

Due to its ubiquity in wireless communications and ease of deployment, we consider the use of orthogonal frequency division multiplexing (OFDM) waveform for localization. In addition to the degrees of freedom that the subcarriers in OFDM provide, we can exploit the multiple antennas at the transmitter and receiver. For simplicity we consider a static uplink setup with a single transmit antenna at a UE and multiple receive antennas at the  BS, i.e. a single-input multiple-output (SIMO) system.
%\footnote{For multiple transmit antennas at the UE, i.e. for a MIMO setup, a UE can perform beamforming to obtain an effective SIMO channel in the uplink.}. 
The uplink SIMO channel estimated at the BS is given by:

\begin{equation}
	\boldsymbol{H} = \left[\boldsymbol{h}_{0}, \boldsymbol{h}_{1}, ..., \boldsymbol{h}_{N_\text{C} -1} \right] \in \mathbb{C}^{N_\text{R} \times N_\text{C}}
	\label{channel}
\end{equation}
where $N_\text{C}$ is the number of subcarriers and $N_\text{R}$ the number of antennas at the receiver. $\boldsymbol{h}_n \in \mathbb{C}^{N_{\text{R}}}$ is the vector describing the CSI for the receive antenna array at  the $n$-th subcarrier. Since $\boldsymbol{H}$ is based on the underlying transfer function between receiver and transmitter, it can be used to obtain a distinct fingerprint for each measured position of a UE.

There exist many techniques to estimate the complex channels based on transmitted pilots. In reality $\boldsymbol{H}$ is the effective channel, i.e. it includes timing offsets and hardware imperfections. Such disturbances may hinder the ability of the raw CSI to provide a distinct fingerprint for each position. 

According to the analysis in\cite{optimumReceiverDesign} and \cite{precisePowerDelayProfiling}, these timing offsets between the oscillators of the transmitter and receiver, influence the estimated channel $\boldsymbol{H}$ and thus its phase at the $k$-th antenna and $n$-th subcarrier, where $k \in [0, N_\text{R}-1]$ and $n \in [0, N_\text{C}-1]$, based on the actual channel $\tilde{\boldsymbol{H}}$ can be expressed as follows:
\begin{equation}
	\angle{\boldsymbol{H}_{n,k}} = \angle{\tilde{\boldsymbol{H}}_{n,k}} + (\tau_\text{p} + \tau_\text{s})n + \tau_\text{c} + \beta + \epsilon_{n,k}
	\label{phase_ofset}
\end{equation}
where the phase of the actual channel at the $k$-th antenna and $n$-th subcarrier is $\angle{\tilde{\boldsymbol{H}}_{n,k}}$. $\tau_{\text{p}}$ is the symbol time offset (STO), $\tau_{\text{s}}$ the sampling time offset, $\tau_{\text{c}}$ the carrier frequency offset, $\beta$ is the phase locked loop (PLL) phase offset and $\epsilon_{n,k}$ is random noise. Due to the continuous timing drift of transmitter and receiver, the estimated channel would not be constant even if the underlying channel does not change. This makes the raw phase practically unusable as a distinct fingerprint for positioning, as pointed out in \cite{wangPhaseCalibration} and \cite{monalisa}.

\subsection{Database Description}

\begin{figure}[t]
	\centering
	\includegraphics[scale=0.3]{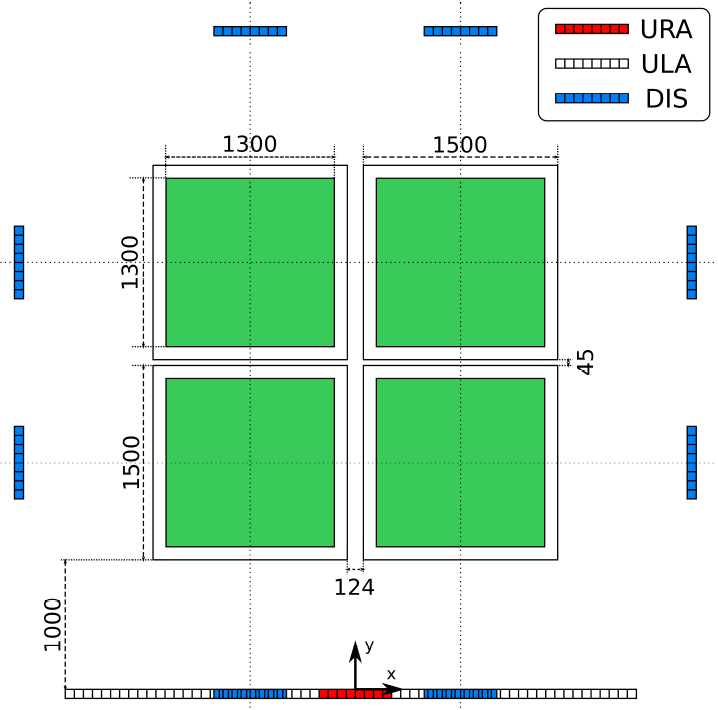}
	\caption{Database measurement scenarios (figure taken from \cite{bast2019csibased}). Distances are indicated in millimeters.}
	\label{KU_Leuven}
\end{figure}

%To evaluate the effectiveness of our proposals we used uplink SIMO channel measurements described in \cite{bast2019csibased} which were obtained using 3 different antenna configurations for the BS in a $2.5 \times 2.5$ m indoor area shown in Fig. \ref{KU_Leuven}. The UE was equiped with a single antenna and moved along a predifined route shown with green. The different antenna configurations of the BS consist of a Uniform Rectangular Array (URA) of $8 \times 8$ antennas, a Uniform Linear Array (ULA) of 64 antennas and a distributed (DIS) configuration of 8 ULA arrays with 8 antennas each. For each configuration, the number of receive antennas at the BS was $N_{\text{R}}=64$. These antennas were used to receive a predefined pilot symbol from the UE at each position.
To evaluate our proposals we use channel measurements which were described in \cite{bast2019csibased} using 3 different antenna configurations for the BS in a $2.5 \times 2.5$ m indoor area shown in Fig. \ref{KU_Leuven}. The different antenna configurations
for the BS consist of a Uniform Rectangular Array (URA) of $8 \times 8$ antennas, a Uniform Linear Array (ULA) of 64 antennas
and a distributed (DIS) configuration of 8 ULA arrays with 8 antennas each. For each configuration, the BS has $N_{\text{R}}=64$ receive antennas. The spacing between
adjacent antenna elements in the ULAs and URA is 70 mm. The UE was equipped with a single antenna. Uplink SIMO channel measurements were performed for equidistantly spaced UE locations (5 mm apart), within the green area inf Fig. \ref{KU_Leuven}. In \cite{bast2019csibased}, the carrier frequency was 2.61 GHz with a bandwidth of 20 MHz and $N_{\text{C}}=100$ subcarriers. 

%In \cite{bast2019csibased}, the carrier frequency was  2.61 GHz with a bandwidth of 20 MHz and $N_{\text{C}}=100$ subcarriers. The spacing between the antenna elements in the ULAs and URA is 70 mm. Channel measurements were performed at uniformly spaced positions, separated by 5 mm.
\begin{figure*}[t]
	\centering
	\begin{subfigure}[t]{0.3\textwidth}
		\includegraphics[scale=0.19]{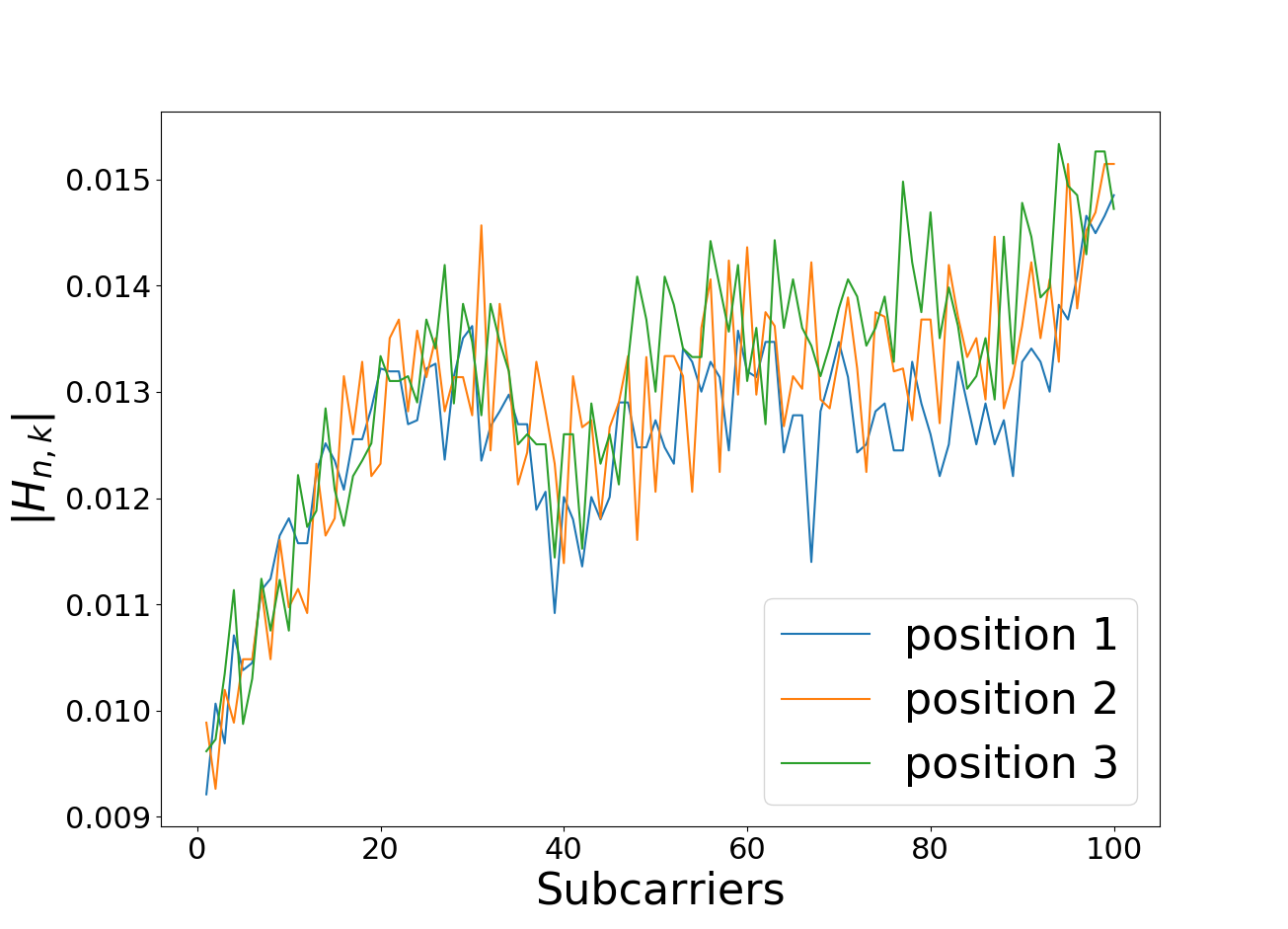}
		\subcaption{Channel magnitude across subcarriers}
		
		\label{magnitude}
	\end{subfigure}\qquad
	\begin{subfigure}[t]{0.3\textwidth}
		\includegraphics[scale=0.19]{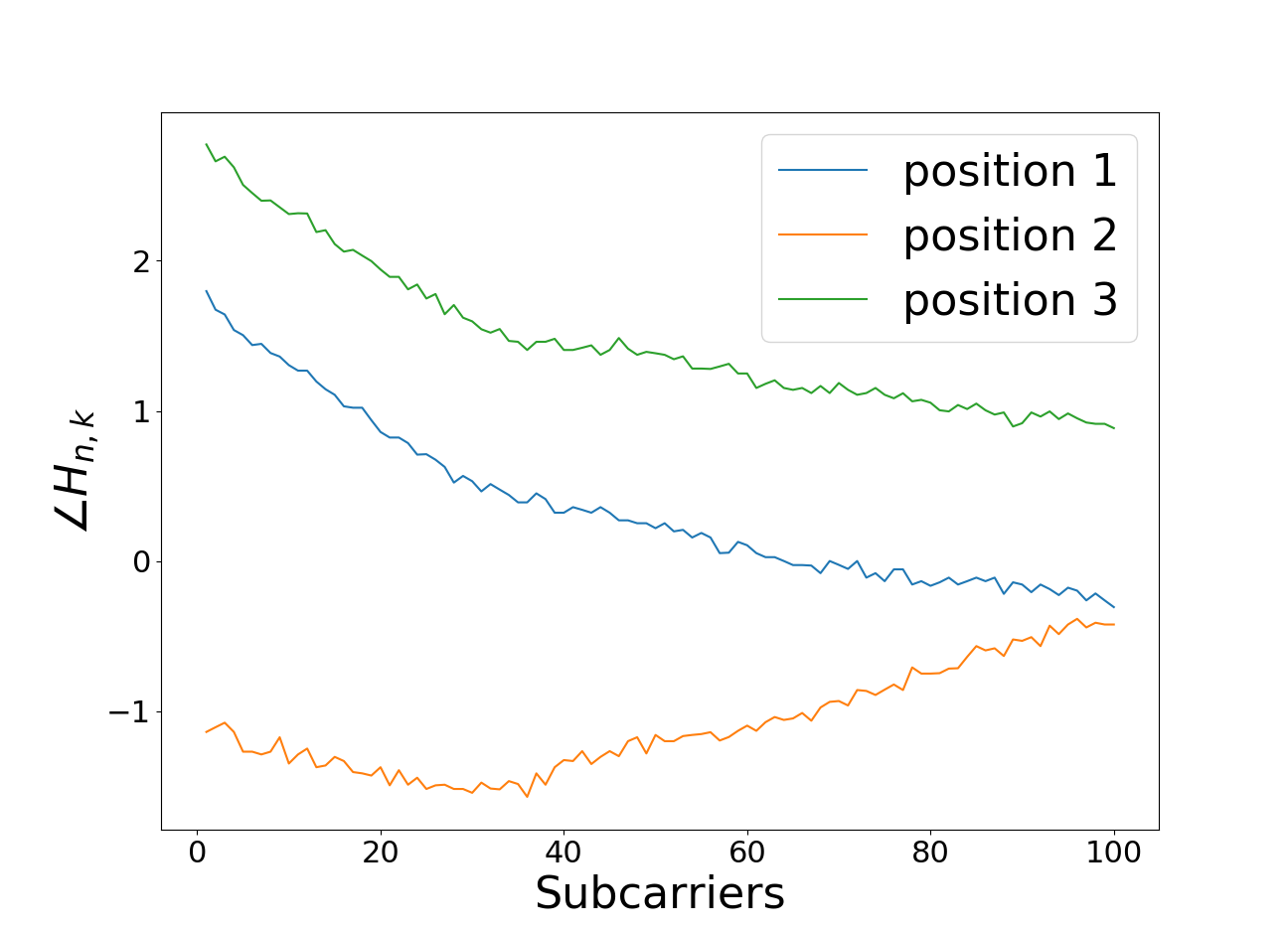}
		\subcaption{Channel phase across subcarriers}
		\label{nopreprocessing}
	\end{subfigure}
	\caption{Channel Fingerprints based on raw CSI}
	\label{Processing}
\end{figure*}
\begin{figure*}[t]
	\centering
	\begin{subfigure}[t]{0.3\textwidth}
		\includegraphics[scale=0.19]{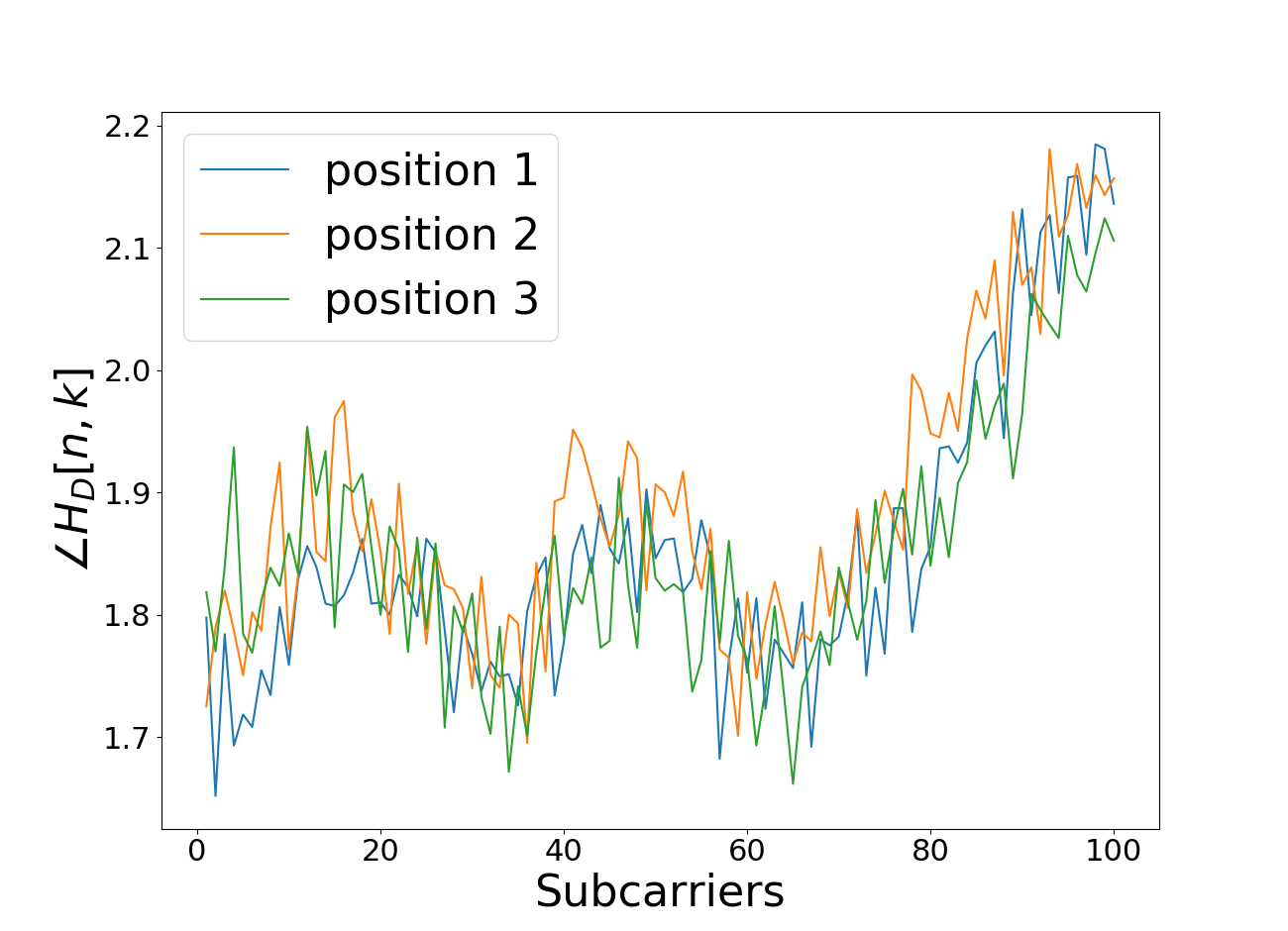}
		\subcaption{Phase difference method across subcarriers}
		\label{difference}		
	\end{subfigure}\qquad
	\begin{subfigure}[t]{0.3\textwidth}
		\includegraphics[scale=0.19]{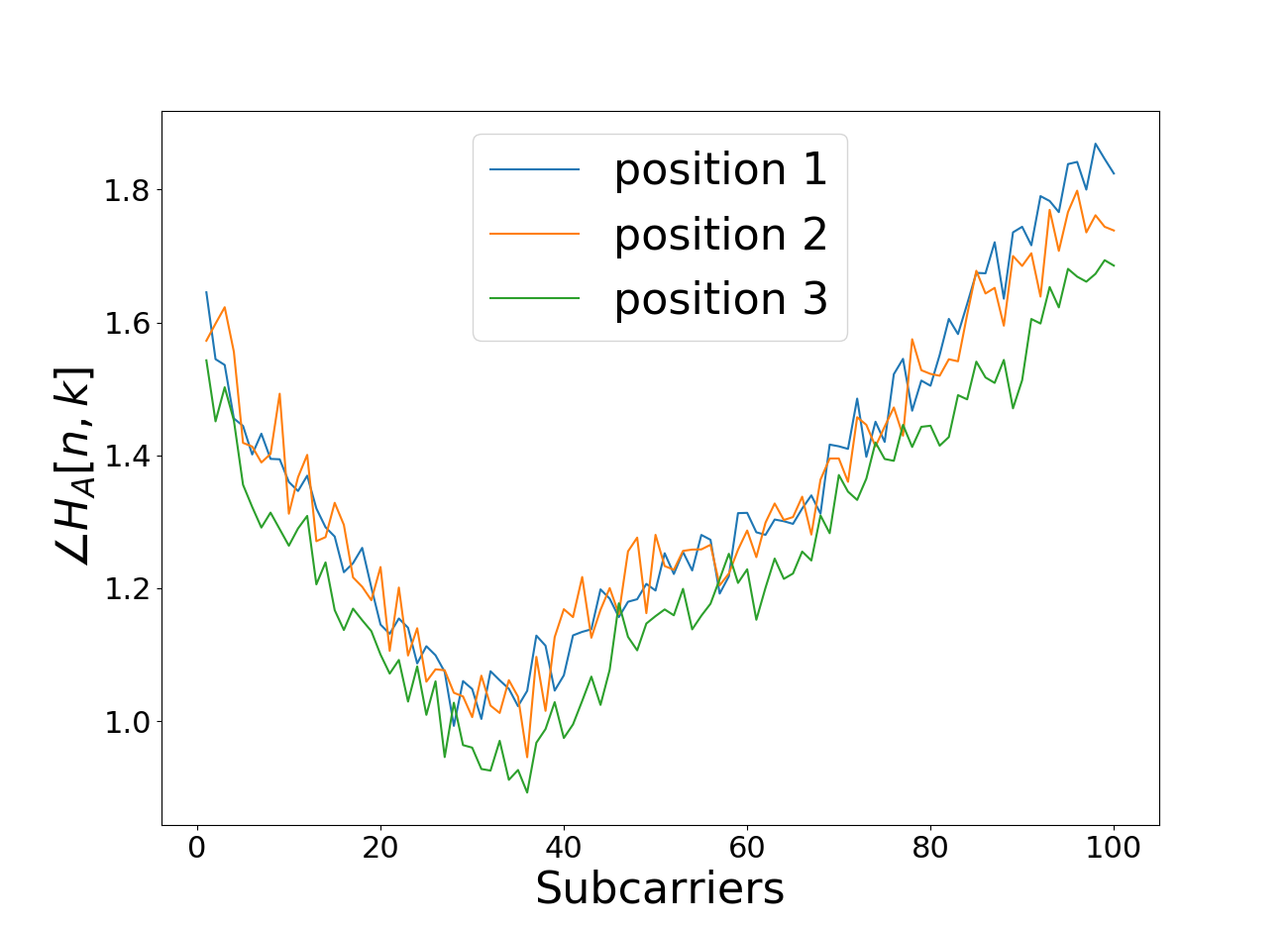}
		\subcaption{Phase alignment method across subcarriers}		
		\label{alignment}
	\end{subfigure}
	\caption{Phase fingerprints based on proposed methods}
	\label{fingerprints}
\end{figure*}
\section{Fingerprint based on Phase Information}

An important aspect for localization based on fingerprint inputs  is that the fingerprint that corresponds to each position is desired to be unique and consistent, during both online and offline phases. Otherwise the correct mapping of a new measurement to the database may not be possible. 

%In particular, the magnitude of the estimated channel is expected to be constant (with some additive noise) and it fulfills the requirements for consistency, as can be seen in Fig. \ref{magnitude}. On the other hand, and although the phase of the underlying channel could also fulfill this requirement,  the real phase of the channel cannot be precisely estimated, and the phase of the effective channel, described in \eqref{phase_ofset}, is unstable. An example of the issue of the phase instability can be seen on Fig. \ref{nopreprocessing} for three different measurements at three neighboring sampled positions (i.e. effectively on the same position). On this figure we see the phase of the estimated channel over the subcarriers for one receiver antenna. Even though all three curves represent three highly correlated measurements, the raw phase estimations are not consistent. 

In particular, the magnitude of the estimated channel can be expected to be spatially consistent, e.g. does not vary significantly (besides some additive noise) over measurements at the same position. The spatial consistency of the magnitude can also be expected for measurements corresponding to very nearby positions. This can be observed in Fig. \ref{magnitude}, where the magnitude over the subcarriers for the first antenna in the ULA is depicted for three neighboring sampled positions of the database from \cite{bast2019csibased}. On the other hand, the raw phase of the estimated channel is usually not spatially consistent as described before. This is shown in Fig. \ref{nopreprocessing}, where the phase over the subcarriers for the first antenna in the ULA is depicted for the same three neighboring positions considered in Fig. \ref{magnitude}.

However, for our purposes we are not interested in estimating the phase of the actual channel $\tilde{\boldsymbol{H}}$, as our goal is merely to acquire a distinct and consistent fingerprint for each position, which is to say for each channel between UE and BS. In the following, we present two methods which produce reliable fingerprints by considering both phase and magnitude, while also preserving the Angle of Arrival (AoA) information embedded in the relationship between the phases of adjacent antennas. The AoA information can be exploited for localization.

%In the following we discuss two ways to achieve that. For both of these methods to work, they not only have to provide a fingerprint, they also have to preserve the information embedded in the phases. In other words, we want to also exploit the information that is provided by the relationship between antennas and not consider each antenna individually, e.g. how AoA is estimated. 

\subsection{Phase Difference}

When the same oscillator is used for all antennas, the phase offsets in \eqref{phase_ofset} are the same for each antenna. Thus, by taking the difference between phases of two adjacent antennas, we eliminate the phase offsets. The phase difference between adjacent antennas for each subcarrier $n$ can be used to produce a distinct fingerprint:

\begin{equation}
	\phi_{n,k} = (\angle \boldsymbol{H}_{n,k} - \angle \boldsymbol{H}_{n,(k+1)\text{mod} N_\text{R}}) +\epsilon 
\end{equation}
where we used the modulo operator to include the phase difference between the last and the first antennas. In this way, we do not lose any information when creating the fingerprints. Based on phase difference, the fingerprint associated with the $k$-th antenna and $n$-th subcarrier for each position would be:

%\begin{equation}
%	\boldsymbol{H}_D[n,k] = |\boldsymbol{H}_{n,k}| \angle \left ( \boldsymbol{H}_{n,k}  \boldsymbol{H}_{n,(k+1)mod N_\text{R}}^* \right ) 
%	\label{Hd}.
%\end{equation}

\begin{equation}
	\boldsymbol{H}_D[n,k] = |\boldsymbol{H}_{n,k}| e^{ \text{j} \left ( \angle \boldsymbol{H}_{n,k}  -\angle \boldsymbol{H}_{n,(k+1)mod N_\text{R}} \right ) }.
	\label{Hd}
\end{equation}

Here we have managed to create a distinct fingerprint $\boldsymbol{H}_D$, for each position, whose phase at each element $[n,k]$ is the phase difference of antennas $k$ and $k+1$ of the original matrix $\boldsymbol{H}$ at subcarrier $n$ while its magnitude is the same as $\boldsymbol{H}$. Also, the difference of phases, takes into account the relationship between antennas. The fingerprint quality offered by the phase difference can be seen in Fig. \ref{difference}, where the phase difference of the first and second antennas in the ULA over the subcarriers is shown for the same positions as in Fig. \ref{nopreprocessing}.
%An example of this method applied, can be seen on Fig. \ref{difference} for the same three neighboring positions as in Fig. \ref{nopreprocessing}. In contrast to Fig. \ref{nopreprocessing}, we can see the fingerprinting quality offered by the processed phase information.

\subsection{Phase Alignment}

\begin{figure*}[t]
	\centering
	\begin{subfigure}[t]{0.3\textwidth}
		\includegraphics[scale=0.19]{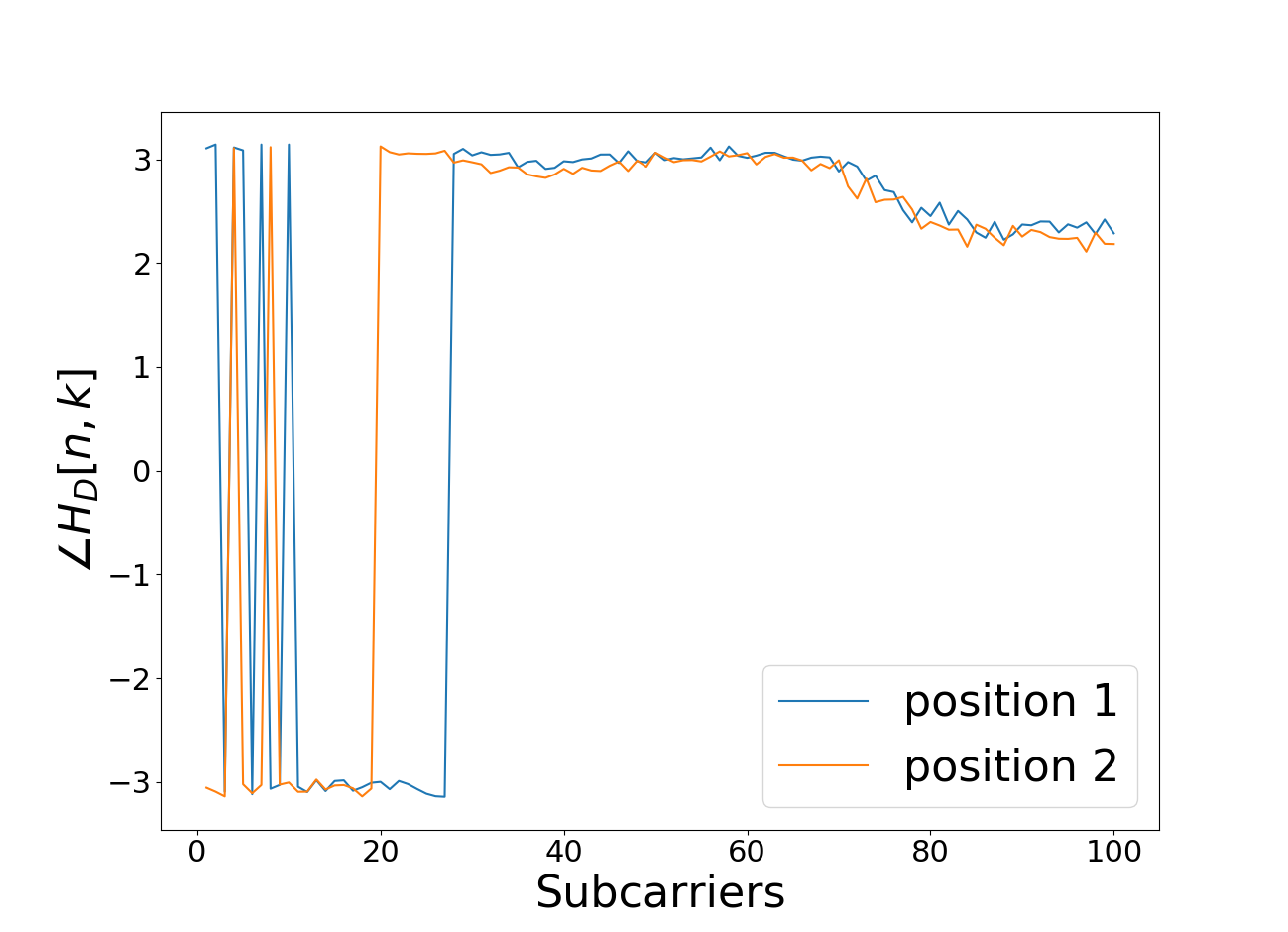}
		\subcaption{Phase wrapping of $\angle H_D[n,k]$}
		\label{wrapping}
	\end{subfigure}
	%\hfill
	\begin{subfigure}[t]{0.3\textwidth}
		\includegraphics[scale=0.19]{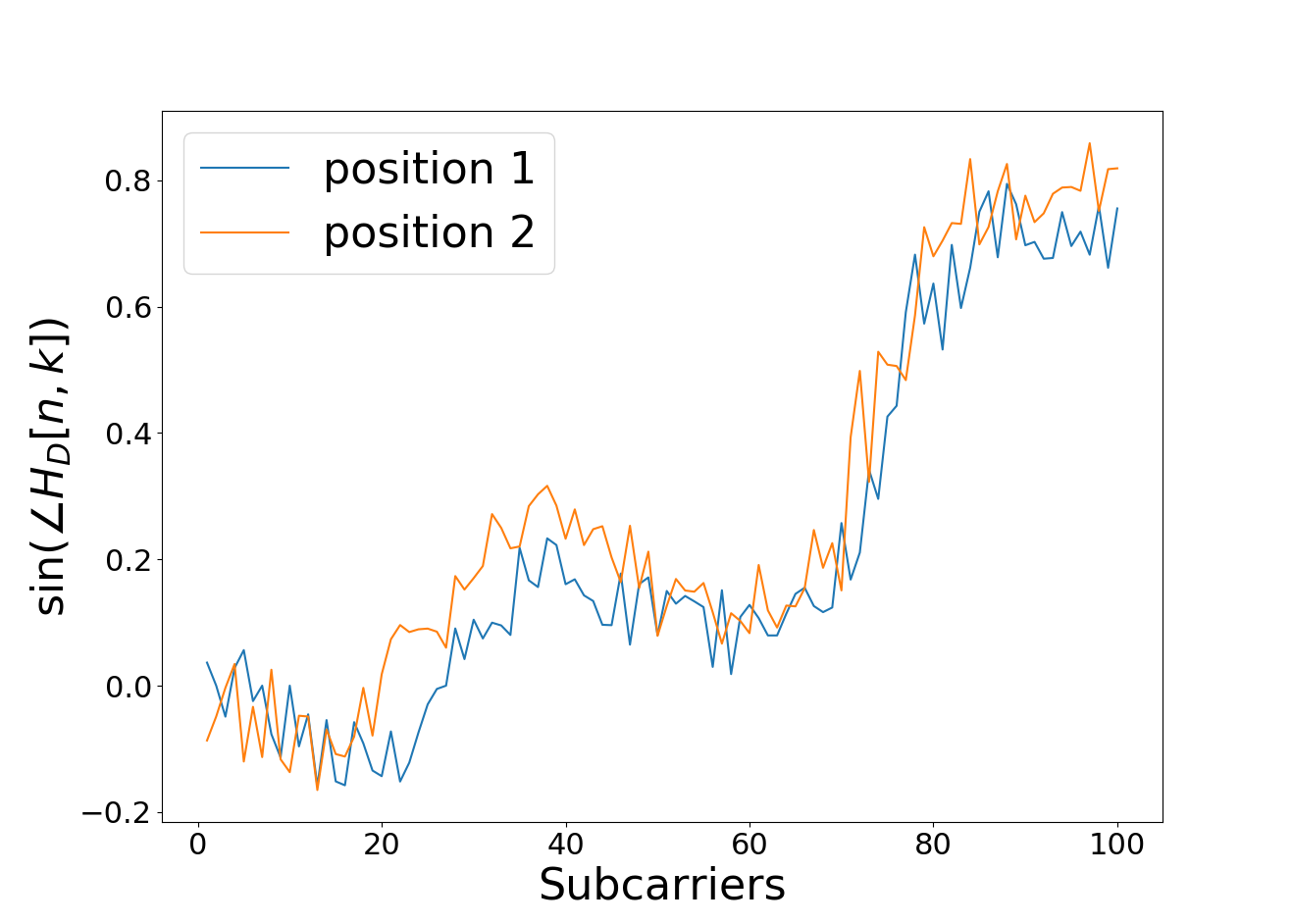}
		\subcaption{Phase Sine}
		\label{sine}
	\end{subfigure}
	%\hfill
	\begin{subfigure}[t]{0.3\textwidth}
		\includegraphics[scale=0.19]{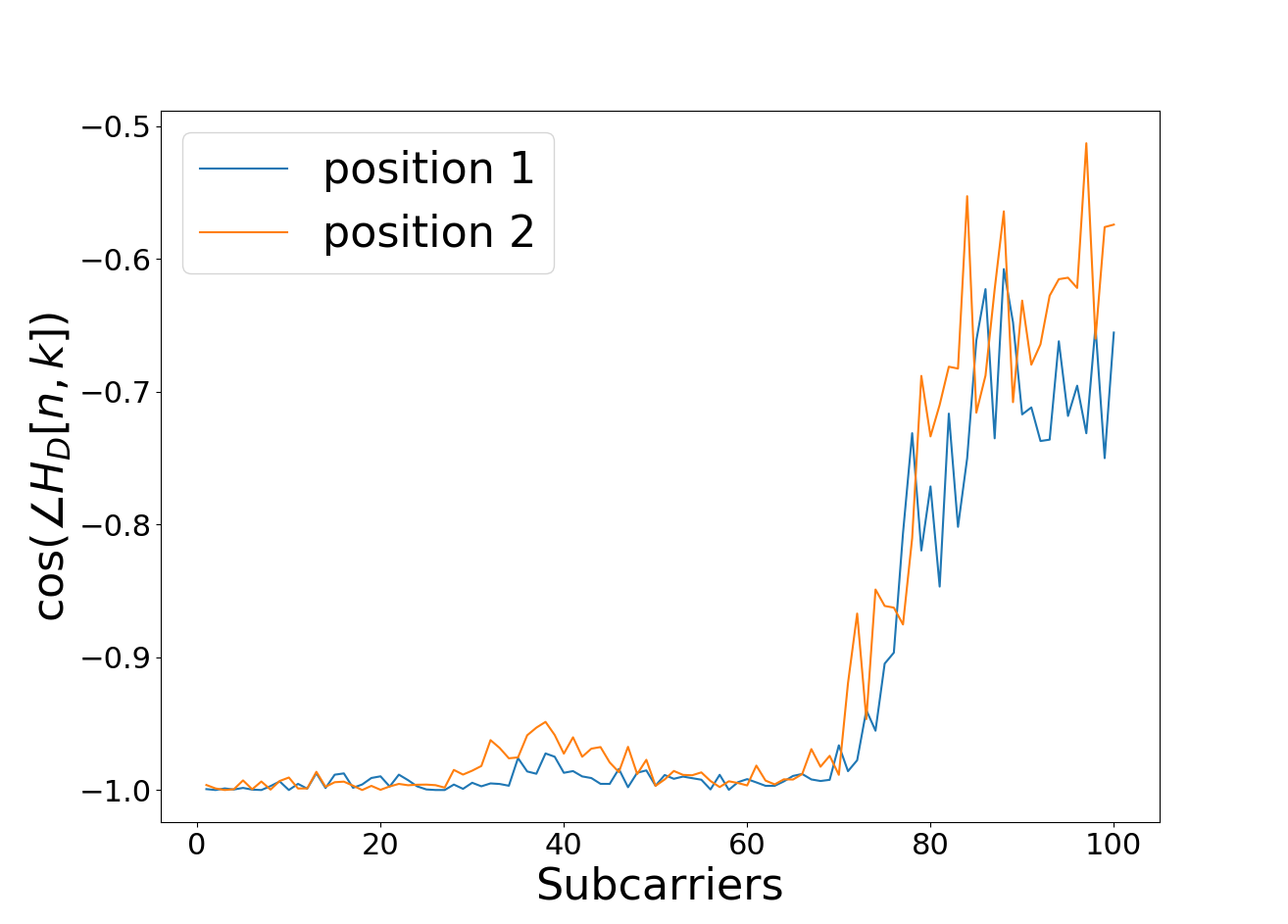}
		\subcaption{Phase Cosine}
		\label{cosine}
	\end{subfigure}

	\caption{Phase fingerprint considering phase wrapping}
	\label{Wrappingprocessing}
\end{figure*}

By considering (3), we see that $\tau_{\text{p}}$ and $\tau_{\text{s}}$ influence the slope of the phases across subcarriers, while $\tau_{\text{c}}$  and $\beta$ simply add a constant offset. These offsets can be mitigated by rotating and shifting the channel across the subcarriers for each antenna as proposed in \cite{monalisa}. However, by applying such a transformation for each antenna separately as suggested in \cite{wangPhaseCalibration}, the AoA information that is embedded in the relationship between the phases of adjacent antennas is lost, i.e., it can not be exploited for localization. To preserve this information, we propose to apply the same transformation across the subcarriers for all antennas. Thus, with this method we obtain the fingerprint $\boldsymbol{H}_A$, where the value associated with the $k$-th antenna and $n$-th subcarrier is calculated as:

\begin{equation}
	\boldsymbol{H}_A[n,k] = \boldsymbol{H}[n,k] e^{-j(\lambda n + b)}
	\label{Ha}
\end{equation}
where $\lambda$ is the reference slope of the subcarrier phases, and $b$ is the reference offset, which are determined as follows.

Firstly, we fit a linear regression model for the phase over the subcarriers of each antenna, resulting in

\begin{equation}
	\angle \boldsymbol{H}[n,k] = \lambda_k n + b_k + \zeta_{n,k}
\end{equation}
where $\zeta_{n,k}$ is the statistical error of the regression model which is minimized. Thus, in contrast to \cite{wangPhaseCalibration} and \cite{monalisa} where the slope $\lambda$ is calculated from the difference of the phases of first and last subcarriers and the offset $b$ as the mean value of the phases across subcarrier, the parameters in \eqref{Ha} are calculated as

\begin{equation}
	\lambda = \frac{1}{N_\text{R}}\sum_{k=1}^{N_\text{R}} \lambda_k, \quad \quad
	b = b_1
\end{equation}
By calculating the slope $\lambda$ this way we avoid the possibility that it will be affected by outliers. As our first method, the proposed phase alignment results in more reliable fingerprints as compared to the raw phase. This can be seen in Fig. \ref{alignment}, where the phase of $\boldsymbol{H}_A$ over the subcarriers associated with the first antenna in the ULA is shown for the same positions considered in Fig. 2b.
%On Fig. \ref{alignment} we see an example of the alignment method. Similar to our first method (Fig. \ref{difference}), the second method is able to provide a distinct and consistent fingerprint of the phase.

\subsection{Phase Wrapping}

Phase wrapping is another problem that can impair the fingerprint. As can be seen in Fig. \ref{wrapping} for a given antenna and two neighboring positions of the database in \cite{bast2019csibased}, phase measurements close to $-\pi$ may fluctuate across the subcarriers, and in some cases the phase wraps around to $\pi$ due to noise. Fig. \ref{wrapping} shows an example of the phase wrapping issue, where the phases across the subcarriers, associated with the first antenna of the ULA of \cite{bast2019csibased}, for two similar positions is shown. While the phase measurements of both positions fluctuate around $-\pi$, they do not wrap around to $\pi$ in the same way, creating two different fingerprints.

The most common method to address this issue is to simply unwrap the phase \cite{wangPhaseCalibration}. This approach, however, is unreliable under noisy conditions as it could lead to large phase values, since the errors accumulate with unwrapping. Such large values then dominate the fingerprint and small variations in the range $[-\pi,\pi)$ will have less influence.

We propose to leverage the fact that for any angle $\theta$ we have $\exp(\text{j}\theta) = \cos(\theta) + \text{j}\sin(\theta)$, such that the information provided by $\theta$ is encoded in $\sin(\theta)$ and $\cos(\theta)$, which are continuous everywhere from $\pi$ to $-\pi$. In Fig. \ref{sine} and \ref{cosine} we plot the sine and cosine of the phase $\angle \boldsymbol{H}_D[n,k]$, in contrast to \ref{wrapping}, we see that the fingerprint quality is preserved when using $\sin{(\angle \boldsymbol{H}_D[n,k])}$ and $\cos{(\angle \boldsymbol{H}_D[n,k])}$. This indicates that the use of real and imaginary parts of the complex valued matrix can solve the wrapping problem since they can be expressed by the magnitude and the cosine and sine of the phase respectively. Additionally, this enables the phase and magnitude to be processed separately, using different and suitable techniques for each one, by using the sine and cosine to represent only the phase information. The downside is that to fully represent the phase fingerprint one must use both those functions, increasing the amount of data to be processed. 

These techniques can also be used on channel estimates based on ray-tracing simulations, even though there are no timing offsets. In this way, the fingerprints from simulations match the fingerprint from measurements, and can be used as an extra layer of information as in \cite{Sousa2018EnhancementOL}.
 
\section{Localization using CNNs}

Although there are several localization schemes based on fingerprint inputs, we focus on using CNNs. For a UE at position $\boldsymbol{r} \in \mathbb{R}^2$, we describe the channel with the function $f$, meaning $f(\boldsymbol{r}) = \boldsymbol{H}_{\star}$, where $\boldsymbol{H}_{\star} \in \{\boldsymbol{H}, \boldsymbol{H}_D, \boldsymbol{H}_A, \boldsymbol{H}_A^\prime \}$ as described in Section II, with $\boldsymbol{H}_A^\prime$ being the fingerprint based on the method proposed in \cite{wangPhaseCalibration}. We will attempt to approximate the inverse function, $f^{-1}(\boldsymbol{H}_\star) = \boldsymbol{r}$, by using CNNs, which have shown promising results for positioning \cite{widmaier2019practical},\cite{Niitsoo2019ADL}. The reason is that CNNs have some features that could be beneficial for the considered type of inputs. 

\subsection{Convolutional Neural Networks}

%\begin{figure}[b]
%	\includegraphics[scale=0.55]{../Images/CNN.png}
%
%	\caption{Convolutional Neural Network}
%	\label{CNN}
%\end{figure}

In a CNN, the input is convoluted with a matrix of smaller dimension, called the kernel. It is almost certainly followed by the pooling operation which is used to reduce the data at the output. Usually, the term CNN is used to describe a NN that uses the convolution operation at some layer.

In addition to the two dimensions (antennas and subcarriers) that our input matrix has, the CNN may also use a third dimension, meaning multiple matrices can be stacked at the input. In machine learning terminology, the input matrices are called channels (not to be confused with the wireless channel). 

This convolutional layer leverages the idea of sparse interactions \cite{Goodfellow-et-al-2016}. A conventional fully connected layer is learning parameters that describe the interactions between each and every one element of the input, while the CNN makes use of the smaller kernel to learn only the interactions between neighboring elements of the input. As the wireless channel between neighboring antennas and subcarriers is usually more correlated than the channels from antennas or subcarriers which are farther apart, the use of CNNs is appealing.

%Although the fingerprint inputs that we proposed in Section III (i.e. \eqref{Hd} and \eqref{Ha}), have a good fingerprint quality as evident from Fig. \ref{fingerprints} and Fig. \ref{Wrappingprocessing}, they are complex valued and need to be transformed into real numbers, in order to be used as input for the CNN. The simplest solution would be to use two different channels of the CNN to provide the magnitude and the phase respectively. For the method proposed to address the phase wrapping, however, one must use three channels of the CNN to fully represent the CSI: one for the magnitude and two for sine and cosine of the phase.
\begin{figure}[!t]
	\centering
	\includegraphics[scale=0.6]{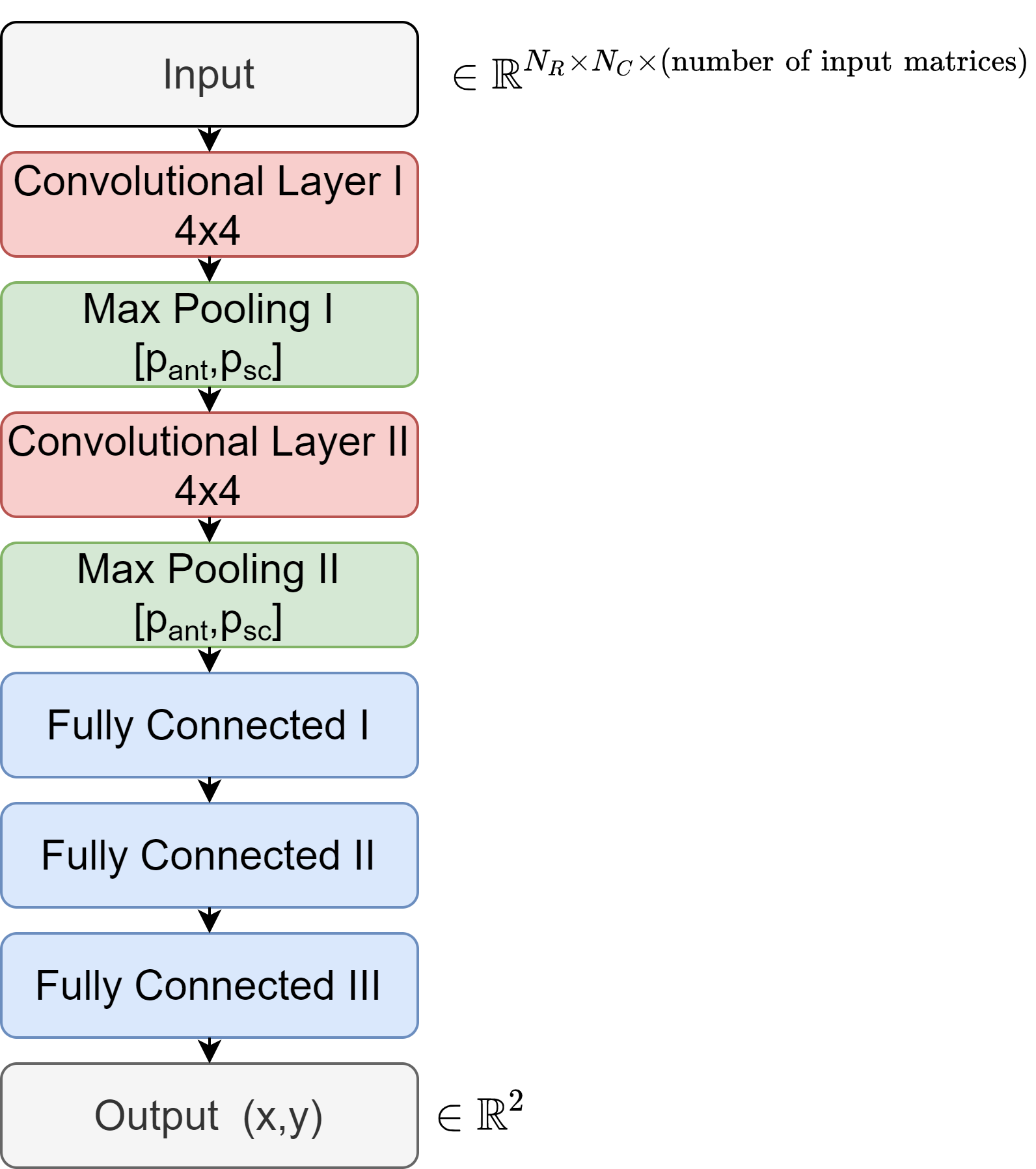}
	
	\caption{Neural Network Model}
	\label{NN_model}
\end{figure}
\subsection{Pooling}

As previously described, after every convolutional layer there is a pooling layer which downsamples the output of the previous layer. The pooling function replaces that output with a summary statistic of the nearby outputs \cite{Goodfellow-et-al-2016}. For example, the max pooling (which we consider in this work) reports the maximum output within a rectangular region of size $p_{ant} \times p_{sc}$, where $p_{ant}$ and $p_{sc}$ are the configurable sizes of the pooling in the antenna and subcarrier dimension, respectively. 
%The two dimensions of this region have to be determined. We refer to $p_1$ for the size of pooling in the antenna dimension and to $p_2$ for the size of pooling in the subcarrier dimension.

Besides reducing the dimension of the data, the pooling layer’s purpose is to make the output invariant to small translations of the input \cite{Goodfellow-et-al-2016}. In our case, the two dimensions of the input matrix are antennas and subcarriers. We expect that small translations in the antenna dimension are important to be detected, since that provides the AoA information. On the other hand adjacent subcarriers within the coherence bandwidth may not provide additional information, as these subcarriers can be correlated. Thus, it may be more beneficial to pool over subcarriers, as pooling over the antennas may lead to a reduction of the angular resolution.

%For the case when the channel is flat, adjacent subcarriers are correlated, so they may not provide additional information. In a realistic indoor scenario, where the channel can be considered to be correlated up to the coherence bandwidth, the optimum pooling dimension may actually depend on the coherence bandwidth of the channel. On the other hand pooling in the antenna dimension may not be so beneficial, , as angular resolution may be reduced.

\section{Simulation Results}

\subsection{Neural Network Setup}

We consider the CNN depicted in Fig. \ref{NN_model} with the input being convoluted with 32 different kernels of dimensions $4 \times 4$. The resulting matrices are pooled, which is followed again by a convolution and pooling layer. The outputs of that layer are vectorized and inputted into four dense layers. The last layer has only 2 neurons expressing the position estimate.

The training set was 80\% of the database of \cite{bast2019csibased} and test set was 20\%. The training procedure starts with a batch size of 32 samples and is increased to 128, 256 and 1024 with each transition set after 30 epochs. The loss function is defined as the Euclidean mean distance of the estimated position and the real position. All the activation functions are set as the Rectified linear unit (ReLU) \cite{Goodfellow-et-al-2016}, except the last one which is linear, and the input data is normalized from 0 to 1, with respect to all the data in the training set. Lastly, we consider as a metric the mean error (ME) given by the Euclidean distance between the estimated and actual position in the test set. 

\subsection{Different Fingerprint Inputs}
\begin{table}[b]
	\caption{One-Channel Input}
	\begin{center}
		\begin{tabular}{|c|c|}
			\hline
			\textbf{Input}& \textbf{ME (m)}\\
			\hline 
			$|\boldsymbol{H}|$ &  0.03805 \\
			\hline
			$\angle \boldsymbol{H}$ &  0.04251 \\
			\hline
			$\angle \boldsymbol{H}_D$ &  0.04088 \\
			\hline
			$\angle \boldsymbol{H}_A$ &  0.03246 \\
			\hline
			$\angle \boldsymbol{H}_A^\prime$ &  0.08142 \\
			\hline
			
		\end{tabular}
		\label{onechannel}
	\end{center}
\end{table}
We first show the results for one, two and three number of input channels considering magnitude and phase information. For all the following configurations the pooling layers had a dimension of $4 \times 4$ ($p_{sc} = p_{ant} = 4$).
We define $|\boldsymbol{H}_{\star}| \in \mathbb{R}^{N_\text{R} \times N_\text{C}}$ and $\angle \boldsymbol{H}_{\star} \in \mathbb{R}^{N_\text{R} \times N_\text{C}} $ as the element-wise absolute value and angle operator, respectively, of the fingerprint matrix $\boldsymbol{H}_{\star} \in \{\boldsymbol{H}, \boldsymbol{H}_D, \boldsymbol{H}_A, \boldsymbol{H}_A^\prime \}$.
%where $\boldsymbol{H}_{\star} \in \{\boldsymbol{H}, \boldsymbol{H}_D, \boldsymbol{H}_A, \boldsymbol{H}_A^\prime \}$ as described in Section II, with $\boldsymbol{H}_A^\prime$ being the fingerprint proposed in \cite{wangPhaseCalibration}. 
For the following results, we consider only the ULA antenna configuration (see Fig. \ref{KU_Leuven}). 

For one input channel of the CNN, Table \ref{onechannel} lists the ME when using the magnitude of the channel, the raw phase and the two different processed phase inputs resulting from the phase difference and phase alignment methods, proposed in Section II. We see that the phase alignment method, not only outperforms using the raw phase, but actually achieves the best performance, even better than using only the magnitude. We also observe that performing a different phase alignment for each antenna as in \cite{wangPhaseCalibration}, deteriorates the performance since the relationship of the phases between antennas (i.e., AoA information) is lost.

Table II presents the results with two input channels, including magnitude and phase as well as the real and imaginary part of the different considered fingerprint matrices, to address the phase wrapping. In addition, we also considered the sine and cosine of $\angle \boldsymbol{H}_D$, thereby using only phase information. Similar to Table \ref{onechannel}, we see that properly processing the phase largely improves the results. We also see that using Re($\boldsymbol{H}_D$) and Im($\boldsymbol{H}_D$) outperforms using the $\sin(\angle \boldsymbol{H}_D)$ and $\cos(\angle \boldsymbol{H}_D)$, as the former includes also magnitude information.

\begin{table}[t]
	\caption{Two-Channel Input}
	\begin{center}
		\begin{tabular}{|c|c|c|}
			\hline
			\textbf{Input I}&\textbf{Input II} &  \textbf{ME (m)}\\
			\hline 
			$|\boldsymbol{H}|$ & $\angle \boldsymbol{H}$ &  0.03126 \\
			\hline
			$|\boldsymbol{H}_D|$ & $\angle \boldsymbol{H}_D$ & 0.03810 \\
			\hline
			$|\boldsymbol{H}_A|$ & $\angle \boldsymbol{H}_A$ & 0.02792 \\
			\hline
			
			$|\boldsymbol{H}_A^\prime|$ & $\angle \boldsymbol{H}_A^\prime$ & 0.03719 \\
			\hline
			Re($\boldsymbol{H}$) & Im($\boldsymbol{H}$) &  0.01809 \\
			\hline
			Re($\boldsymbol{H}_A$) & Im($\boldsymbol{H}_A$) &  0.01614 \\
			\hline
			Re($\boldsymbol{H}_D$) & Im($\boldsymbol{H}_D$) &  0.01316 \\
			\hline
			Re($\boldsymbol{H}_A^\prime$) & Im($\boldsymbol{H}_A^\prime$) &  0.03478 \\
			\hline
			sin($\angle \boldsymbol{H}_D$)& cos($\angle \boldsymbol{H}_D$) &  0.01425 \\
			\hline
		\end{tabular}
		\label{twochannel}
	\end{center}
\end{table}

Lastly, Table \ref{threechannel} provides results with three channel inputs, where we can use the magnitude of the channel with the sine and cosine of the phase of the considered fingerprints. As in previous results, the use of properly processed phase information provides the best performance. From both Table \ref{twochannel} and \ref{threechannel} we see that using the fingerprints based on matrix $\boldsymbol{H}_D$ while also addressing phase wrapping achieved the best performance. The small improvement when using three channels can be attributed to the fact the the CNN is able to employ different processing for phase and magnitude, and extract the relevant information in each case.
%Comparing the last row of Table \ref{twochannel} and \ref{threechannel}, we note that considering a third input channel for the magnitude leads to a ME reduction of $\sim 9\%$.

%Tastly, we will provide the results for when we use three channel inputs. By using three inputs, we are able to use the two phase inputs sine and cosine to avoid the unwrapping, as well as the magnitude of the channel. These are presented in Table \ref{threechannel}. As expected, when we use all the available information, processed accordingly, we get the best performance. This reinforces the proposal that a properly processed phase of the channel should be taken into account for localization applications.

\begin{table}[t]
	\centering
	\caption{Three-Channel Input}
	
	\begin{center}
		\begin{tabular}{|c|c|c|c|}
			\hline
			\textbf{Input I}&\textbf{Input II}&\textbf{Input III}&  \textbf{ME (m)}\\
			\hline 
			$|\boldsymbol{H}|$ & sin($\angle \boldsymbol{H}$)& cos($\angle \boldsymbol{H}$) & 0.01981 \\
			\hline
			$|\boldsymbol{H}_A|$ & sin($\angle \boldsymbol{H}_A$)& cos($\angle \boldsymbol{H}_A$) & 0.01734 \\
			\hline
			$|\boldsymbol{H}_D|$ & sin($\angle \boldsymbol{H}_D$)& cos($\angle \boldsymbol{H}_D$) & 0.01290 \\
			\hline
			
		\end{tabular}
		\label{threechannel}
	\end{center}
\end{table}

\subsection{Pooling}

In this subsection, we analyze the impact of different pooling options $[p_{ant},p_{sc}]$ on the positioning performance, considering the ULA, URA and DIS antenna configurations from \cite{bast2019csibased}. For the evaluation, we use $|\boldsymbol{H}_D|$, $\sin(\angle \boldsymbol{H}_D)$ and $\cos(\angle \boldsymbol{H}_D)$ as the input channels of the CNN shown in Fig. \ref{NN_model}. In Table IV, we show the ME for different pooling options with $p_{ant} \times p_{sc} =4$, such that that resulting CNNs have the same complexity. For each antenna configuration, pooling over the subcarriers, i.e. [1,4], leads to the smallest ME, while the largest ME is obtained when pooling over the antennas, i.e. [4,1].  For a given pooling option, the best performance is achieved with the distributed antenna configuration, as it collects CSI at distinct locations around a UE’s position. On the other hand, the worst performance results by using the URA, since its resolution on the horizontal plane where the UE lies, is smaller compared to the other antenna configurations.

The results in Table IV suggest that it is more beneficial to pool over the subcarriers than over the antennas. Thus, in Fig. 6 we examine the ME with pooling option $[1,p_{sc}]$ for different values of $p_{sc}$. For each antenna configuration, we see there is an optimum pooling $p_{sc}$ over the subcarriers, which we posit that it depends on the coherence bandwidth of the channel. 

The lowest ME attained for each of the ULA, DIST and URA antenna configurations  are 6.11 mm, 5.20 mm and 9.81 mm respectively, which is lower than the ones reported in \cite{bast2019csibased}. This was achieved by using the optimal size of pooling $p_{sc}$, which is different for each antenna configuration, and the three channel input: $|\boldsymbol{H}_D|,\sin(\angle \boldsymbol{H}_D), \cos(\angle \boldsymbol{H}_D)$.

\begin{table}[t]
	\caption{Different Pooling Options}
	\begin{center}
		\begin{tabular}{|c|c|c|}
			\hline
			\textbf{Pooling [$p_{ant}$, $p_{sc}$]}&\textbf{Antenna Configuration} & \textbf{ME (m)}\\
			\hline 
			
			[1,4] & ULA & 0.00612 \\
			\hline
			[2,2] & ULA & 0.01010  \\
			\hline
			[4,1] & ULA & 0.01633 \\
			\hline
			[1,4] & distributed & 0.00521 \\
			\hline
			[2,2] & distributed &0.00732  \\
			\hline
			[4,1] & distributed & 0.00982 \\
			\hline
			[1,4] & URA & 0.01096 \\
			\hline
			[2,2] & URA & 0.01183  \\
			\hline
			[4,1] & URA & 0.01908 \\
			\hline
		\end{tabular}
		\label{pooling_table}
	\end{center}
\end{table}
\begin{figure}[t]
	\centering
	\includegraphics[scale=0.45]{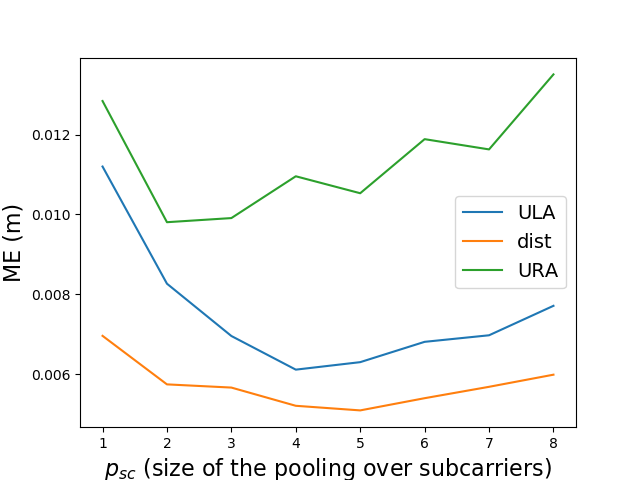}
	
	\caption{ME for different pooling dimensions}
	\label{pooling}
\end{figure}
\section{Conclusion}

We have examined the use of CSI over multiple antennas and subcarriers, as fingerprint inputs of a CNN for UE localization. As the raw phase of the estimated channel cannot be used as a consistent fingerprint, we have presented different methods for producing reliable fingerprints based on phase information. Although the proposed methods have been evaluated with CNNs, they can also be used for other localization schemes based on fingerprints. For different number of inputs of a CNN, simulation results have shown that UE localization can be improved with properly processed phase information. We have also investigated the impact of different pooling options on the positioning performance with CNNs, showing that it is more beneficial to pool over the subcarriers than over the antennas. Simulation results have shown there is an optimum pooling size over the subcarriers, whose dependency on the coherence bandwidth is part of future work. 

%In addition, we have investigated the impact of different pooling options on the positioning performance. Simulation results indicate that pooling over the subcarriers is more beneficial than pooling over the antennas. We have also shown that there is an optimum size of the pooling over the subcarriers, which we believe may depend on the coherence bandwidth of the channel. In the future we plan to investigate further the relationship between coherence bandwidth and the pooling size, as well as the impact of different antenna configurations and the use of deeper CNNs.

%We examine the problem of UE localization using channel measurements as fingerprints and investigate the properties a fingerprint should have. We conclude that the raw phase of the complex channel should be preprocessed in order to achieve those properties. Additionally we show that the CNN is a suitable tool for this kind of problem and examine its structure more closely and its relation to the wireless channel. We believe that other neural networks, particularly deeper ones, could perform better than the one presented here, nevertheless our analysis holds. Namely that the phase of the channel holds important information that can be extracted and also that the coherence bandwidth of the channel should inform the structure of the neural network, particularly with regards to pooling dimension. In the future we plan to investigate the relationship between coherence bandwidth and pooling. 

\bibliographystyle{IEEEtran}
\bibliography{references}

\end{document}